\documentstyle[12pt]{article}
\begin{document}
\tolerance=5000
\def\be{\begin{equation}}
\def\ee{\end{equation}}
\def\bea{\begin{eqnarray}}
\def\eea{\end{eqnarray}}
\def\nn{\nonumber \\}
\def\cF{{\cal F}}
\def\det{{\rm det\,}}
\def\Tr{{\rm Tr\,}}
\def\e{{\rm e}}
\def\tr{{\rm tr\,}}

\  \hfill 
\begin{minipage}{3.5cm}
OCHA-PP-104 \\
NDA-FP-37 \\
September 1997 \\
\end{minipage}

\ 

\vfill

\begin{center}

{\large\bf Vacuum polarization 
of supersymmetric D-brane \\
in the constant electromagnetic field}

\vfill

{\sc Tomoko KADOYOSHI}\footnote{e-mail: 
kado@fs.cc.ocha.ac.jp}, 
{\sc Shin'ichi NOJIRI$^{\clubsuit}$}\footnote{
e-mail : nojiri@cc.nda.ac.jp}, \\
{\sc Sergei D. ODINTSOV$^{\spadesuit}$}\footnote{
e-mail : odintsov@quantum.univalle.edu.co, 
odintsov@kakuri2-pc.phys.sci.hiroshima-u.ac.jp}
and 
{\sc Akio SUGAMOTO}\footnote{
e-mail: sugamoto@phys.ocha.ac.jp}

\vfill

{\sl Department of Physics, 
Ochanomizu University \\
Otsuka, Bunkyou-ku Tokyo 112, JAPAN}

\ 

{\sl $\clubsuit$ 
Department of Mathematics and Physics \\
National Defence Academy, 
Hashirimizu Yokosuka 239, JAPAN}

\ 

{\sl $\spadesuit$ 
Tomsk Pedagogical University, 634041 Tomsk, RUSSIA \\
and \\
Dep.de Fisica, Universidad del Valle, 
AA25360, Cali, COLOMBIA \\
}

\vfill

{\bf ABSTRACT}

\end{center}

We study the vacuum polarization of supersymmetric 
toroidal D-brane placed in the constant electromagnetic field. 
Explicit calculation of the one-loop effective 
potential is performed for membrane with constant magnetic 
or electric background. We find the one-loop 
potentials vanish as the effect of supersymmetry, 

\ 


\newpage

D-brane theory \cite{2,3,4} may be considered as 
one of the most essential ingredients in the study of $D=11$ 
M-theory \cite{1} which plays the role of new theory 
of everything. 
Therefore, recently there was much interest in the investigation 
of D-branes \cite{5} and their supersymmetric 
generalizations \cite{Ab,So,Ag}.

In the attempts to understand quantum properties of 
such objects, the vacuum polarization effects may be 
quite important.

Towards this purpose, the vacuum energy (or effective 
potential) for bosonic p-branes on the toroidal or 
spherical background has been calculated \cite{8,9}. 
(Such study has been extended for toroidal D-brane 
on constant electromagnetic field in ref.\cite{NO}). 
It is known that the toroidal background 
in $D=11$ supermembrane \cite{DI} is supersymmetric 
one and the vacuum energy is zero.

If the supersymmetric toroidal D-branes is placed 
in the constant electomagnetic background, 
however, the situation might be 
different and the vacuum energy could not vanish 
because not for all supersymmetric backgrounds effective
potential is zero.
 Note that all 
supersymmetries are conserved on such background, cf.\cite{14}. 

The purpose of the present letter is to explicitly study the 
effective potential for supersymmetric toroidal 
D-brane \cite{Ab} in the constant electromagnetic 
background.
We will find that the vacuum energy vanishes as in the 
usual toroidal D-brane.

The supersymmetric D-brane is described by the following 
Dirac-Born-Infeld-type action (see Refs.\cite{Ab,So,Ag})
\be
\label{DBIS0}
S_D=k\int_0^T d\zeta_0 \int d^p\zeta \e^{-\phi(X)}
\left[\det \left(G_{MN}
\partial_iZ^M \partial_jZ^N +F_{ij}\right)
\right]^{1 \over 2} + S_{WZ}\ .
\ee
Here $\zeta^i$'s are the coordinates on the D-brane world 
sheet ($i,j=0,1,\cdots, p$), 
$F_{ij}$ is the electromagnetic field strength on 
the D-brane world sheet:
\be
\label{fstrgth}
F_{ij}=\partial_i A_j - \partial_j A_i \ ,
\ee
and $Z^M$ are the superspace 
coordinates of D-brane
\be
\label{Z}
\{Z^M\}=(X^\mu, \theta^\alpha)
\ee
($\mu, \nu = 0,1,\cdots,9$, $\alpha=1,2,\cdots,32$).
 $\theta$ is a Majorana-Weyl (16 independent 
components) spinor in case of heterotic string, 
Majorana spinor in case of type IIA superstring 
and two kinds of Majorana-Weyl (totally 32 components) 
spinors in case of type IIB superstring.
The $\phi(X)$ is dilaton field and 
$S_{WZ}$ is Wess-Zumino like term containing 
anti-symmetric tensor fields.

We now consider the case  when the action is 
given by (without WZ term which is not essential in our context)
\be
\label{DBIS}
S_D=k\int_0^T d\zeta_0 \int d^p\zeta \e^{-\phi(X)}
\left[\det \left((G_{\mu\nu}+B_{\mu\nu})
\Pi_i^\mu \Pi_j^\mu +F_{ij}\right)
\right]^{1 \over 2}\ .
\ee
$G_{\mu\nu}$ is the metric of the space-time, 
$B_{\mu\nu}$ is the anti-symmetric tensor 
and 
\be
\Pi_i^\mu \equiv \partial_i X^\mu
- i\bar\theta \Gamma^\mu \partial_i \theta\ ,
\ee
where $\Gamma^\mu=e_a^\mu\gamma^a$ ($e_a^\mu$ is the vielbein 
field and $\gamma^a$'s denote $\gamma$ matrices in 10 
dimensions).  
For simplicity, we assume $G_{\mu\nu}$ and $B_{\mu\nu}$ are 
constant.
The action (\ref{DBIS}) is invariant under the
global super-Poincar\'e transformation:
\bea
\label{sP}
\delta X^\mu &=& i\bar\epsilon \Gamma^\mu \theta \nn
\delta \theta &=& \epsilon
\eea 
The correspondent transformations for vector and tensor fields 
are not written explicitly.
The system also has 
local reparametrization invariance with respect to 
$\zeta$ and $\kappa$-symmetry, 
the latter is a fermionic 
symmetry given by
\be
\label{kappa}
\delta\theta=(1+\Gamma)\kappa\ .
\ee
Here $\kappa$ is a spinor parameter and $\Gamma$ is a 
matrix which satisfies
\be
\label{Gamma}
\tr \Gamma = 0\ ,\ \ \Gamma^2=1\ .
\ee 
Note that correspondent transformations for bosonic components 
and vectors are not written explicitly.
The reparametrization invariance is fixed by
choosing the gauge condition
\be
\label{ggch}
X^i=R_i \zeta^i \ \ (R_0=1)\ \ \ \ i=0,1,\cdots,p\ .
\ee
The $\kappa$ symmetry is also fixed by choosing
\be
\label{kappagf}
\Gamma \theta= - \theta\ .
\ee
Note that in the above gauge, there appear no 
Faddeev-Popov ghosts.

We are now interesting in the stability of  
D-brane when $F_{ij}$ (or 
$B_{ij}$) has nontrivial vacuum expectation value
and study the effective potential, which is defined by
\be
\label{EP}
V=-\lim_{T\rightarrow\infty}{1 \over T}
\ln \int {DX^\bot D\theta^\alpha DA_i \over V_A}\e^{-S_D}\ .
\ee
Here $V_A$ is the gauge volume for the gauge field $A_i$ and 
\be
\label{Xbot}
X^\bot = (X^{p+1},X^{p+2},\cdots,X^9) \ .
\ee
The integration in Eq.(\ref{EP}) with respect to $\theta$ 
should be understood to be integrated over 
the $\theta^\alpha$ space restricted by (\ref{kappagf}).
We now also impose the periodic boundary 
conditions corresponding to the toroidal D-brane
\bea
\label{pbc}
&& X^m(\zeta^0, \zeta^1, \zeta^2, \cdots, \zeta^p)
 =X^m(\zeta^0+T, \zeta^1, \zeta^2, \cdots, \zeta^p) \nn
&& =X^m(\zeta^0, \zeta^1+1, \zeta^2, \cdots, \zeta^p) 
 =X^m(\zeta^0, \zeta^1, \zeta^2+1, \cdots, \zeta^p) \nn
&& = \cdots =X^m(\zeta^0, \zeta^1, \zeta^2, \cdots, \zeta^p+1) \nn
&& (m=p+1,p+2,\cdots,9)\ ,\nn
&& \theta^\alpha(\zeta^0, \zeta^1, \zeta^2, \cdots, \zeta^p)
 =\theta^\alpha(\zeta^0+T, \zeta^1, \zeta^2, \cdots, \zeta^p) \nn
&& =\theta^\alpha(\zeta^0, \zeta^1+1, \zeta^2, \cdots, \zeta^p) 
 =\theta^\alpha(\zeta^0, \zeta^1, \zeta^2+1, \cdots, \zeta^p) \nn
&& = \cdots =\theta^\alpha(\zeta^0, \zeta^1, \zeta^2, \cdots, \zeta^p+1)\ ,\\
&& (\alpha=1,2,\cdots, 16 \ \mbox{for Type IIA, IIB 
superstring}, 8\ \mbox{for heterotic string} \nonumber
\eea
For simplicity, we only consider here the 
Type IIA and IIB superstrings with 
\bea
\label{smpl}
&& \phi=0 \nn
&& G_{\mu\nu}=\delta_{\mu\nu} \nn
&& B_{mn}=B_{im}=B_{mi}=0\ , \hskip 1cm 
(i=0,1,\cdots,p;\ m,n=p+1, \cdots, 9)
\eea
and assume that anti-symmetric tensor fields in $S_{WZ}$ vanish.
We also divide the anti-symmetric part $\cF_{ij}$ in 
$\hat G_{ij}$ into the sum of 
the classical part $\cF_{ij}^c$ and 
the quantum fluctuation $F_{ij}^q$:
\be
\label{div}
\cF_{ij}=\cF_{ij}^c + F_{ij}^q\ ,
\ee

In the following, we write $R_i^2\delta_{ij}+\cF_{ij}^c$ 
as $\hat G_{ij}$ and $F_{ij}^q$ as $F_{ij}$.

Then expanding the action up to the quadratic 
terms with respect to  quantum fields, we obtain
\bea
\label{quad}
S_2&=&\int_0^T d\zeta_0 \left[
(\det \hat G_{ij})^{1 \over 2}+  
\int d^p\zeta \left\{ {1 \over 2} 
G^{Sij}\partial_i X^\bot
\cdot\partial_j X^\bot-i(\hat G^{-1})^{ij}\bar\theta\Gamma_i 
\partial_j \theta\right. \right. \nn
&& + \left.\left.
\left( -{1 \over 2}(\hat G^{-1})^{jk}(\hat G^{-1})^{li}
+{1 \over 4}(\hat G^{-1})^{ji}(\hat G^{-1})^{lk} 
\right)F_{ij}F_{kl} \right\}\right]
\eea
Here
\be
\label{Gtensor}
G^{Sij}\equiv{1 \over 2} 
\left((\hat G^{-1})^{ij}+(\hat G^{-1})^{ji}
\right) \ .
\ee
is defined with the inverse matrix $(\hat G^{-1})^{ij}$ 
of $\hat G_{ij}$.
Since the contribution to the one-loop effective 
potential from 
the $X^\bot$ and the gauge fields  was calculated 
in \cite{NO}, we now consider the contribution 
$V_\theta$ from $\theta$, which may be found as follows: 
\bea
\label{thetaV}
V_\theta &=&-{1 \over 4}
\Tr\ln (\hat G^{-1})^{ij}\Gamma_i\partial_j \nn
&=&-{1 \over 8}\Tr\ln \left( (\hat G^{-1})^{ij}
\Gamma_i\partial_j 
\right)^2 \nn
&=&-4\Tr\ln g_{ik} (\hat G^{-1})^{ij}(\hat G^{-1})^{kl}
\partial_j \partial_l\ .
\eea
Here we have used
\bea
\label{Gamma2}
\{ \Gamma_i, \Gamma_j\}&=&2g_{ij} \nn
g_{ij}&=&R_i^2 \delta_{ij}\ .
\eea
Then if we define the following matrices
\bea
\label{Gtheta}
\hat G_\theta^{ij} &\equiv& g_{kl} (\hat G^{-1})^{li}
(\hat G^{-1})^{kj} \nn
\tilde G_\theta^{\alpha\beta} &=& 
\hat G_\theta^{\alpha\beta} -  
{1 \over \hat G_\theta^{00}} 
\hat G_\theta^{0\alpha} \hat G_\theta^{0\beta}
\eea
we find 
\be
\label{Vtheta2}
V_\theta=-4\sum_{n_1,n_2,\cdots , n_p=-\infty}^\infty
\left( 4\pi^2\sum_{\alpha,\beta=1}^p 
\tilde G_\theta^{\alpha\beta} n_\alpha n_\beta
\right)^{1 \over 2}\ .
\ee

In the following, we consider some examples for 
the choice of electromagnetic background.

First we consider the membrane ($p=2$) with 
the magnetic background
\be
\label{mag}
\cF^c_{0k}=0\ ,\ \ \cF^c_{12}=-\cF^c_{21}=h 
\ee
and assume
\be
\label{R}
R_1=R_2=R\ .
\ee
Then we obtain
\bea
\label{dete}
\hat G&\equiv& \det \hat G_{ij}= R^4 + h^2 \\
\label{inv}
(\hat G^{-1})^{ij}&=&
\left(
\begin{array}{ccc}
1 & 0 & 0 \\
0 & \hat G^{-1}R^2 & -\hat G^{-1}h \\
0 & \hat G^{-1}h & \hat G^{-1}R^2 
\end{array} \right) \ ,
\eea
and we find 
\be
\label{Vthetam}
V_\theta = -{8 R \over (R^4+ h^2)^{1 \over 2}}f_T(1,1)\ .
\ee
Since the bosonic contribution is given by \cite{NO}
\be
\label{Vb}
V_b = {8 R \over (R^4+ h^2)^{1 \over 2}}f_T(1,1)\ ,
\ee
the quantum corrections in the one-loop potential cancel with
each other and the total one-loop correction vanishes.
Here we have used zeta-function regularization (see ref.\cite{E}
for a review) and obtain 
\be
\label{fT}
f_T(1,1)=2\pi\sum_{n_1,n_2=-\infty}^\infty \left(
n_1^2+n_2^2 \right)^{1 \over 2}=-1.438\cdots\ .
\ee

We also consider the one-loop effective 
potential of the membrane 
in the constant electric background where
\be
\label{ele}
\cF_{0k}=e\ ,\ \ \cF_{kl}=0\ ,\ \ R_1=R_2=R\ .
\ee
Then we find
\bea
\label{detG}
\hat G&\equiv& \det \hat G_{ij}=R^4 + 2e^2 R^2 \\
\label{inv2}
(\hat G^{-1})^{ij}&=& \hat G^{-1}
\left(
\begin{array}{ccc}
R^4 & -eR^2 & -eR^2 \\
eR^2 & R^2+e^2 & -e^2 \\
eR^2 & -e^2 & R^2+e^2 
\end{array}
\right) \ .
\eea
Then the contribution from $\theta$ is given by 
\be
\label{Vthetae}
V_\theta=-{4(R^2 + e^2)^{1 \over 2} 
\over R^2} \hat f_T(1,1, -{e^2 \over R^2 + e^2})\ .
\ee
The contribution is exactly cancelled by the 
bosonic contribution found in ref.\cite{NO}.
Here $\hat f_T(1,1, s)$ is defined by using of modified Bessel 
function $K_\nu$ as follows,
\bea
\label{hatf}
\hat f_T(1,1, s)&\equiv& 2\pi \sum_{n_1,n_2=-\infty}^\infty
\left(n_1^2 + n_2^2 + 2s n_1 n_2 \right)^{1 \over 2} \nn
&=&4\pi\zeta(-1) - {\zeta(3) \over 2\pi} \Delta \nn
&& - 4 \Delta^{1 \over 2} \sum_{n=1}^\infty
{\sigma_2(n)\cos (2sn\pi) \over n}
K_1(\pi n \Delta^{1 \over 2})\ , 
\eea
and 
\be
\label{Delt}
\Delta\equiv 4-4s^2\ ,\ \ \sigma_s(n)\equiv \sum_{d|n} d^s\ .
\ee
Hence, by direct calculation of one-loop effective potential 
for supersymmetric D-brane on supersymmetric electromagnetic background 
we showed that such effective potential is zero. It would be interesting 
to extend the results of our work for other electromagnetic
backgrounds without supersymmetry. The stability of such backgrounds 
maybe studied using above results.

\noindent
Acknoweledgments. We would like to thank the referee of this paper  
for pointing out the mistake which completely changed the final result 
of our work.

\end{document}